# Evidence of Time-dependent Diffusive Shock Acceleration in the 2022 September 5 Solar Energetic Particle Event

Xiaohang Chen[1], Lulu Zhao[1], Joe Giacalone[2], Nishtha Sachdeva[1], Igor V. Sokolov[1], Gábor Tóth[1], Christina M. S. Cohen[3], David Lario[4], Fan Guo[5], Athanasios Kouloumvakos[6], Tamas I. Gombosi[1], Zhenguang Huang[1], Ward B. Manchester, IV[1], Bart van der Holst[1], Weihao Liu[1], David J. McComas[7], Matthew E. Hill[6], and George C. Ho[8]

[1] Department of Climate and Space Sciences and Engineering, University of Michigan, Ann Arbor, MI 48109, USA; zhlulu@umich.edu
[2] Lunar and Planetary Laboratory, University of Arizona, Tucson, AZ 85721, USA
[3] California Institute of Technology, Pasadena, CA 91125, USA
[4] NASA Goddard Space Flight Center, Greenbelt, MD 20771, USA
[5] Los Alamos National Laboratory, Los Alamos, NM 87545, USA
[6] Applied Physics Laboratory, Johns Hopkins University, Laurel, MD 20723, USA
[7] Department of Astrophysical Sciences, Princeton University, Princeton, NJ 08544, USA
[8] Southwest Research Institute, San Antonio, TX 78238, USA

## Abstract

On 2022 September 5, a large solar energetic particle (SEP) event was detected by Parker Solar Probe (PSP) and Solar Orbiter (SolO) at heliocentric distances of 0.07 and 0.71 au, respectively. PSP observed an unusual velocity dispersion signature: particles below ∼1 MeV exhibited a normal velocity dispersion, while higher-energy particles displayed an inverse velocity arrival (IVA) feature, with the most energetic particles arriving later than those at lower energies. The maximum energy increased from about 20–30 MeV upstream to over 60 MeV downstream of the shock. The arrival of SEPs at PSP was significantly delayed relative to the expected onset of the eruption. In contrast, SolO detected a typical large SEP event characterized by a regular velocity dispersion at all energies up to 100 MeV. To understand these features, we simulate particle acceleration and transport from the shock to the observers with our newly developed SEP model—Particle ARizona and MIchigan Solver on Advected Nodes. Our results reveal that the IVA and delayed particle onset detected by PSP originate from the time-dependent diffusive shock acceleration processes. After shock passage, PSP's magnetic connectivity gradually shifted due to its high velocity near perihelion, detecting high-energy SEPs streaming sunward. Conversely, SolO maintained a stable magnetic connection to the strong shock region where efficient acceleration was achieved. These results underscore the importance of spatial and temporal dependence in SEP acceleration at interplanetary shocks and provide new insights to understand SEP variations in the inner heliosphere.

## 1. Introduction

Solar energetic particles (SEPs) are a key aspect of space weather because they can pose serious radiation hazards to humans and electronic equipment in space (see reviews by D. V. Reames 1999; M. Desai & J. Giacalone 2016 and references therein). Large SEP events are associated with fast and wide coronal mass ejections (CMEs) and strong shocks (e.g., S. W. Kahler et al. 1984; E. W. Cliver et al. 2004; R. Chandra et al. 2013; E. W. Cliver & E. D'Huys 2018; A. Kouloumvakos et al. 2019). Significant longitudinal variations in SEP intensities are usually observed at 1 au from multiple spacecraft observations (H. V. Cane et al. 1986; D. V. Reames et al. 1996, 2013; D. Lario et al. 2016). These variations are believed to be related to a number of factors regarding the properties of the CME-driven shock and background solar wind, such as magnetic connectivity, CME speed, shock geometry, and cross-field diffusion (F. Guo et al. 2010; J. Giacalone 2012, 2017; X. Kong et al. 2017; D. Lario et al. 2017; X. Chen et al. 2022; W. Liu et al. 2023; N. Wijsen et al. 2023). During shock arrival, there can be an increase in the intensities of lower-energy SEPs (less than a few MeV), known as energetic storm particle events (D. A. Bryant et al. 1962). It is believed that these particles are accelerated locally by the shock, providing valuable insights into the mechanisms of particle acceleration and the properties of shock waves. In contrast, high-energy particles (>10 MeV) have a peak intensity that is generally detected well before the shock arrival, with intensities gradually declining after reaching a peak. These observations suggest that most high-energy SEPs are accelerated in the regions near the Sun where the magnetic field strength and turbulence are strong (e.g., M. B. Kallenrode et al. 1993; W. K. M. Rice et al. 2003). As the shock expands outward, the acceleration rates diminish, lowering the maximum energy to which particles can be accelerated. Consequently, by the time the shock arrives at Earth, it rarely accelerates particles beyond 10 MeV nucleon$^{-1}$.

It is now widely accepted that diffusive shock acceleration (DSA) is the primary mechanism for accelerating CME-related SEPs (W. I. Axford et al. 1977; G. F. Krymskii 1977; A. R. Bell 1978; R. D. Blandford & J. P. Ostriker 1978; J. R. Jokipii 1982). Charged particles are scattered by magnetic irregularities in the solar wind and accelerated rapidly as they traverse the strong compression across the shock multiple times. DSA inherently predicts a power-law momentum distribution $f \propto p^{-\gamma}$ with the spectral index $\gamma$ only depending on the density compression ratio. Note that this arises from the assumption of an infinite, steady, and planar shock. The single power-law spectrum would break if introducing other physical processes, such as back-reaction, finite acceleration time, or shock curvature (J. F. McKenzie & H. J. Voelk 1982; P. Blasi 2002; B. B. Wang et al. 2022). In space- and time-dependent scenarios, the duration $\tau_A$ required to accelerate a particle at a planar shock from an initial momentum $p_0$ to a final momentum $p$ can be described as (L. O. Drury 1983, 1991; M. A. Forman & L. O. Drury 1983; J. R. Jokipii 1987)

$$\tau_A = \int_{p_0}^{p} \frac{dp'}{p'} \frac{3}{U_1 - U_2} \int_{-\infty}^{\infty} dx' \exp\left(-\left| \int_0^{x'} \frac{U}{\kappa(x, p)} dx \right|\right), \tag{1}$$

where $U_1$ and $U_2$ are the upstream and downstream flow speeds normal to the shock front, measured near the shock in the shock frame of reference (where the shock is at $x = 0$); $U$ is the plasma flow speed; and $\kappa(x, p)$ is the spatial- and momentum-dependent component of the diffusion tensor along the shock normal. Equation (1) reveals that the acceleration time increases with the value of the diffusion coefficient. Given a certain amount of time $\tau_A$, particles can be accelerated to higher energies with a smaller $\kappa$. Consequently, shock geometry can substantially modulate acceleration efficiency as $\kappa_\perp \ll \kappa_\parallel$, and particles are accelerated more rapidly in the region near the Sun of stronger magnetic turbulence (smaller $\kappa$). However, CME-driven shocks are neither steady-state nor planar as they propagate through the interplanetary magnetic field (IMF). The acceleration rate can vary significantly in both space and time. Therefore, understanding the processes underlying SEP acceleration and transport near the Sun is crucial for interpreting SEP observations at 1 au and thus improving the reliability of space weather forecasts.

The launch of recent missions, such as Parker Solar Probe (PSP; N. J. Fox et al. 2016) and Solar Orbiter (SolO; D. Müller et al. 2020), facilitates an in-depth exploration of the inner heliosphere at an unprecedented distance. The minimum perihelion distances of PSP and SolO can approach as close as 0.0478 and 0.28 au from the center of the Sun, respectively. This proximity provides a unique opportunity to investigate SEP acceleration and transport in the early stages. On 2022 September 5, a large SEP event was observed by PSP at 15.45 $R_\odot$ just before its perihelion of orbit 13. This event is associated with an extremely fast CME ($>$2200 km s$^{-1}$) erupting from the far side of the Sun relative to Earth (E. Paouris et al. 2023). SolO also observed the event at a distance of 0.71 au, with a longitudinal separation of $\sim$32° westward from PSP (Figure 1). Notably, this is one of the most intense SEP events detected by SolO to date, with particle energies reaching beyond 100 MeV nucleon$^{-1}$. The CME and associated SEPs have been carefully analyzed by a number of studies. D. M. Long et al. (2023) studied the magnetic configuration of the CME by combining remote sensing and in situ observations and showed that the flux rope formed along the magnetic inversion line beneath the heliospheric current sheet (HCS). The evolution of the CME and associated solar wind structures was examined by E. Paouris et al. (2023) and O. M. Romeo et al. (2023). Hard X-rays (HXRs) and gamma rays were also observed during the eruption (J. G. Mitchell et al. 2024; J. T. Vievering et al. 2024). D. Trotta et al. (2024) examined the interplanetary shock properties and nearby turbulence, revealing more disturbed upstream magnetic field conditions at SolO compared to PSP. C. M. S. Cohen et al. (2024) provided an overview of SEP observations from PSP, highlighting uncommon phenomena such as delayed onsets of high-energy SEPs and highly anisotropic sunward flows. To understand these features, we employ Particle ARizona and MIchigan Solver on Advected Nodes (PARMISAN)—a data-driven, physics-based model—to simulate the SEP observations of PSP and SolO. We find that the variations in the SEP time–intensity profiles are highly dependent on the characteristics of the shock regions that observers are magnetically connected to. We also report the evidence of time-dependent DSA at the initial stage of the event observed by PSP. These results highlight how the evolution of SEP acceleration and transport near the Sun contributes to the spatial variations observed at distant heliospheric locations.

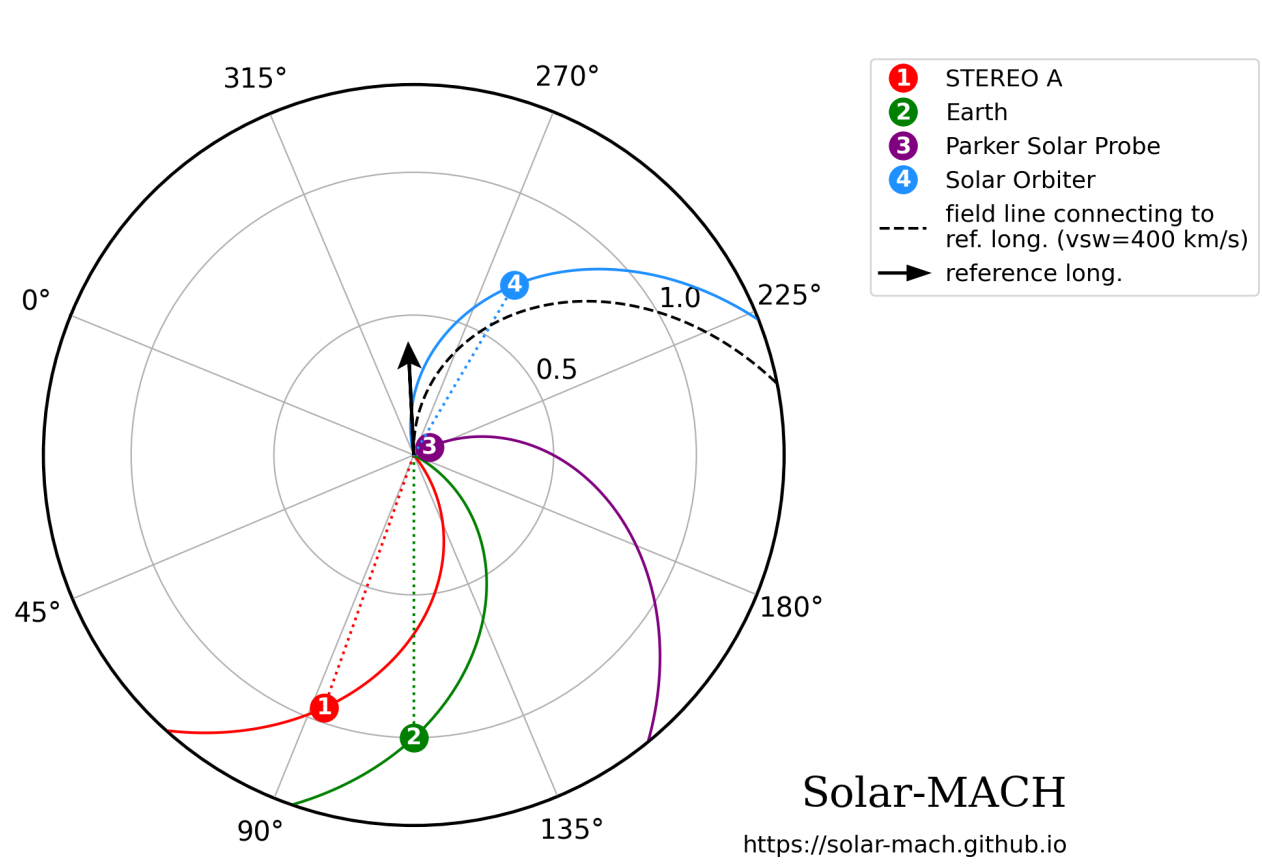

**Figure 1.** Longitudinal spacecraft constellation and magnetic connectivity at 16:00 UT on 2022 September 5, generated with the Solar-MACH tool (J. Gieseler et al. 2023). The colored markers indicate the locations of STEREO-A (red), Earth (green), PSP (purple), and SolO (blue). The black arrow denotes the longitude of the parent active region. The colored lines and dashed black line from the active region represent the nominal Parker spirals of 400 km s$^{-1}$ solar wind.

The structure of this paper is as follows. The overview of the SEP event observed by PSP and SolO on 2022 September 5 is described in Section 2. We introduce the components of the SEP model in Section 3. In Section 4, we discuss the simulation results and compare them with the in situ observations. Finally, the conclusions are provided in Section 5.

## 2. Observations

The solar event on 2022 September 5 originated from the active region (NOAA AR 13088) on the far side of the Sun as seen from Earth (E. Paouris et al. 2023). During this period, SolO was optimally positioned to observe the active region. An M9-class flare was observed in HXR by the Spectrometer Telescope for Imaging X-rays (S. Krucker et al. 2020) on board SolO (J. T. Vievering et al. 2024). While near-Earth observations did not capture the eruption, the coronagraphic images from the Sun Earth Connection Coronal and Heliospheric Investigation (R. A. Howard et al. 2008) on board STEREO-A provided high-cadence (5 minutes) observations of the CME and its associated shock evolution in the corona. The detailed CME structures was also captured by the Wide-Field Imager coronagraph (A. Vourlidas et al. 2016) on board PSP. The leading-edge velocity of the CME was estimated to exceed 2200 km s$^{-1}$ (E. Paouris et al. 2023; R. Patel et al. 2023), marking it as one of the fastest CMEs observed in solar cycle 25 to date.

A large SEP event associated with this eruption was observed by both PSP and SolO. As the eruption occurred on the solar far side relative to Earth, near-Earth observations are not considered in this study. Figure 1 presents the relative positions of multiple spacecraft and Earth at 16:00 UT on 2022 September 5, where the circles mark the observer locations (STEREO-A, red; Earth, green; PSP, purple; SolO, blue), the solid lines trace the nominal Parker spirals connecting each observer with the Sun, and the black arrow indicates the eruption longitude where the CME originated. PSP detected the onset of SEPs at around 16:45 UT and crossed the east flank of CME-driven shock at a heliocentric distance of only 15.45 $R_\odot$, just prior to the perihelion of orbit 13. The SEP observations were collected by the Integrated Science Investigation of the Sun (IS⊙IS) on board PSP (D. J. McComas et al. 2016). IS⊙IS includes two energetic particle instruments (EPIs) with overlapping energy ranges: EPI-Lo, optimized for lower energies, and EPI-Hi, designed for higher energies. EPI-Lo consists of 80 small apertures, with fields of view covering nearly an entire hemisphere. It measures particles using the time-of-flight versus energy technique and determines the composition, spectra, and anisotropies of particles with energies from ∼20 keV to several MeV nucleon$^{-1}$ (M. E. Hill et al. 2017). EPI-Hi includes the low- and high-energy telescopes (LETs and HETs), which employ the standard $dE/dx$ versus $E$ measurement method (M. E. Wiedenbeck et al. 2017). LETs detect ions with energies from around 1 to tens of MeV nucleon$^{-1}$, while HETs measure ions from about 10 to over 100 MeV nucleon$^{-1}$ and electrons with energies between 1 and 6 MeV. The LET-A aperture is oriented 45° west of the Sun–spacecraft line, i.e., along the nominal Parker spiral direction at 1 au. The HET-A aperture points 20° west of the Sun relative to the spacecraft–Sun line. LET-B and HET-B are aligned oppositely to LET-A and HET-A, respectively, and LET-C is oriented orthogonally to both LET-A and LET-B.

Figure 2 provides an overview of the SEP event observed by PSP on 2022 September 5. Figure 2(a) displays the magnetic field magnitude and components in the RTN coordinate system (FIELDS; S. D. Bale et al. 2016). The heliocentric distance of PSP is shown along the top $x$-axis in solar radii. The vertical dotted line marks the beginning of the CME eruption around 16:10 UT, followed by the arrival of the CME-driven shock at 17:27:20 UT (O. M. Romeo et al. 2023). Figures 2(b)–(d) present the energy spectrograms for proton intensities from HET-A, LET-A, and EPI-Lo, respectively. Note that the spectrogram of EPI-Lo is made using all looking directions. Figure 2(e) shows the EPI-Lo measurements as a function of $1/v$, where $v$ is the particle speed. Due to elevated particle intensities, EPI-Hi activated dynamic threshold modes shortly after the onset of SEPs, and EPI-Lo shut down approximately at the time of shock passage (see details in C. M. S. Cohen et al. 2024). The gray bars in the spectrograms indicate periods of missing data during the event. One of the most notable features is the onset profile of SEPs in the spectrograms. A velocity dispersion for energies below ∼1 MeV can be determined by fitting the first-arriving SEPs in the $1/v$ plot, as indicated by the gray dashed curves in Figures 2(d) and (e). The first-arriving particles are assumed to have propagated without undergoing scattering processes; thus, higher-energy particles reach the observer earlier than lower-energy ones. However, in this event, the onset of the high-energy population (above 1 MeV) occurs later than that of the lower-energy components. This observation indicates a delay in the release of high-energy SEPs from the shock, outlined by the red dashed curve in Figure 2(d), and is also detected in the EPI-Hi data in Figures 2(b) and (c). Note that this feature has been observed upstream of Earth's bow shock, known as "inverse velocity dispersion," which is attributed to an extended connection time between the spacecraft and the bow shock (e.g., G. C. Anagnostopoulos et al. 1986; E. T. Sarris et al. 1987). In contrast, our analysis reveals that the delayed arrival of high-energy particles in this event is primarily caused by a longer acceleration timescale at the CME-driven shock, as we demonstrate in the following sections. To distinguish this acceleration-driven effect from the transport-related phenomena, we propose using the term "inverse velocity arrival" (IVA; following Z. Xu et al. 2024) rather than "inverse velocity dispersion" to describe this observation. Several numerical studies have attempted to explain this feature by invoking a later magnetic connection to the supercritical shock region (A. Kouloumvakos et al. 2025), particles escaping near the shock front (Z. Ding et al. 2024; T. M. Do et al. 2025) or the evolution of the shock and connectivity (Z. Ding et al. 2025; Y. Li et al. 2025). These interpretations require a small acceleration rate at the beginning of the eruption, which is unusual given the extreme CME speed in this event. Therefore, it is essential to resolve the 3D CME structure and investigate how it modulates SEP acceleration and transport during the CME expansion.

The initial solar particle release (SPR) time can be inferred from the velocity dispersion analysis by assuming that ions of all species and energies are released from the source regions simultaneously (e.g., D. V. Reames 2009; T. Laitinen et al. 2015). In this event, however, the estimated SPR time of the first particles detected by PSP is approximately 16:45 UT (Figure 2(e)), significantly later than the CME eruption around 16:10 UT. This delay may be explained either by assuming that particle acceleration occurred not at the time of the solar eruption but later as the shock forms at higher altitudes or by considering the delayed timing at which the observer becomes magnetically connected to the shock surface (S. Kahler 1994). Another possibility is that the discrepancy results from particles traveling a longer path along the magnetic field to reach the observer (e.g., R. Chhiber et al. 2021). However, considering the CME's extreme speed and proximity to the Sun, none of these hypotheses fully accounts for the observed delay of over 35 minutes.

The maximum energy of the SEPs ($E_{\mathrm{Max}}$) observed upstream of the shock is about 20 MeV. Then, the intensity in high energies suddenly decreases after the shock crossing (Figure 2(b)). This reduction feature is also detected in heavy ions (C–Fe), which are unaffected by the dynamic threshold modes. Bidirectional electron fluxes are observed during this period, suggesting a possible flux rope crossing (O. M. Romeo et al. 2023). The intensities recovered after 18:00 UT in the downstream and $E_{\mathrm{Max}}$ increased to over 40 MeV. C. M. S. Cohen et al. (2024) analyzed the "pixel" counting rates from EPI-Hi, which do not have thresholds applied, and found most of the SEPs moving toward the Sun (highly anti-sunward anisotropy) during the phase of increasing intensity.

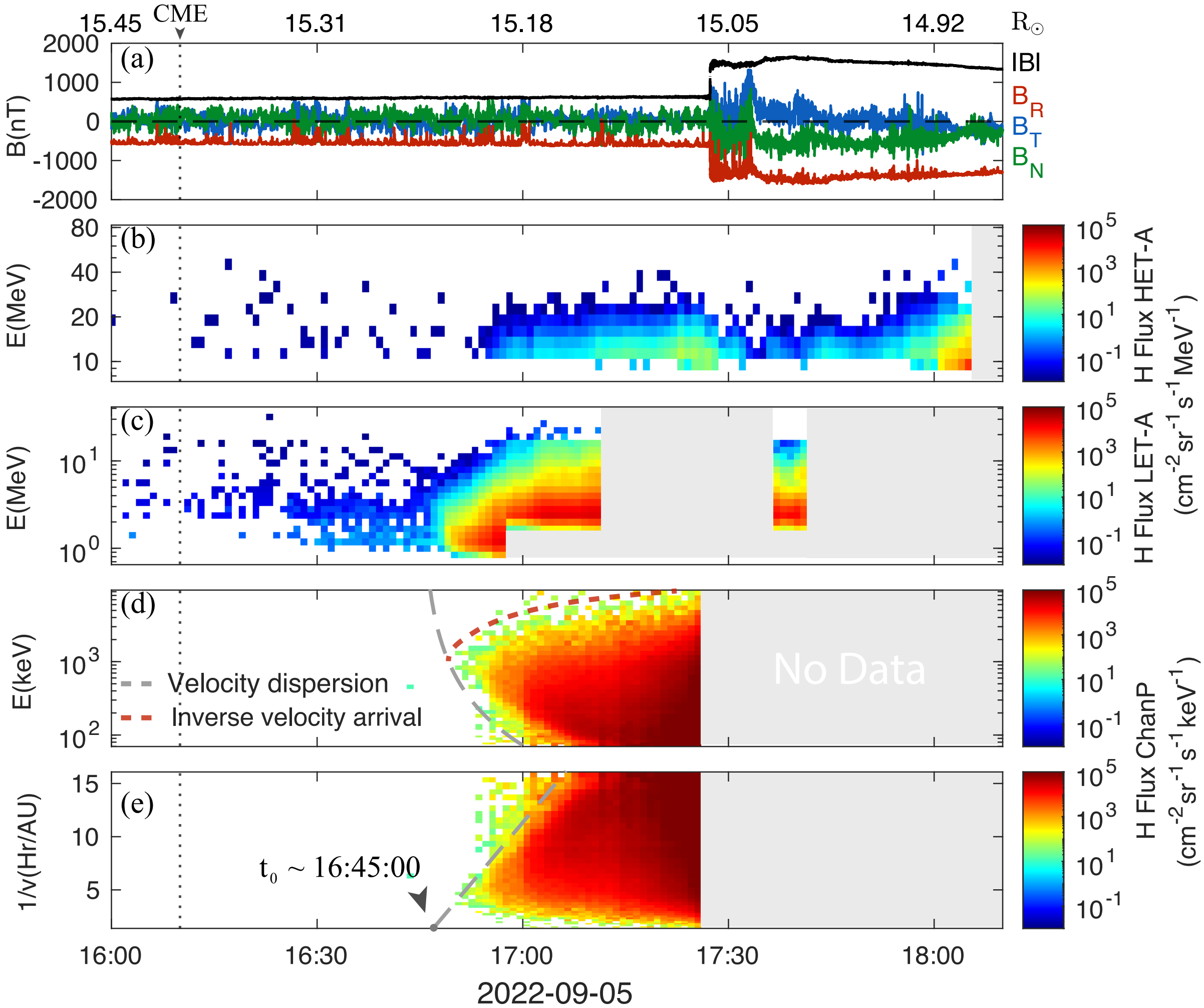

**Figure 2.** Overview of the PSP observations on 2022 September 5 from 16:00 to 18:15 UT. (a) Magnetic field magnitude and components in the RTN coordinate system, with the heliocentric distance of PSP shown along the top *x*-axis in solar radii. Energy spectrograms for proton intensities from (b) HET-A, (c) LET-A, and (d) EPI-Lo of all the looking directions. (e) Energy spectrograms for EPI-Lo plotted against inverse velocity. The vertical dotted line indicates the CME eruption time. The gray and red dashed lines represent the regular velocity dispersion and IVA, respectively.

This SEP event was also well observed by SolO, positioned at approximately 0.71 au and about 32° west of PSP. Figure 3 displays the particle and field observations from SolO covering the time interval of 2022 September 5 00:00 UT to September 7 10:00 UT. The top panel displays the magnetic field data (T. S. Horbury et al. 2020) in RTN coordinates. On the top of Figure 3, we indicate the heliocentric distance of the spacecraft, similar to Figure 2. The SEP observations were obtained from the Energetic Particle Detector (EPD; J. Rodríguez-Pacheco et al. 2020) suite of instruments. In particular, we use the Suprathermal Electrons and Protons (STEP) telescope, which measures low-energy (~2–80 keV) ions with a field of view roughly along the nominal Parker spiral direction; the sunward aperture of the Electron Proton Telescope (EPT), which measures ions between 25 keV and 6.4 MeV; and the sunward aperture of the HET, which measures protons from ~6.8 to 107 MeV. Figure 3(b) shows the differential ion flux, multiplied by the square of ion energy, from STEP, sunward EPT, and sunward HET, covering ion energies from 10 keV nucleon$^{-1}$ to 100 MeV nucleon$^{-1}$. The SEP fluxes are plotted as a $1/v$ function in Figure 3(c). Gray bars indicate periods of low-quality data due to particle "overflow," resulting from the high intensities during the event. High-energy SEPs were detected well before the shock arrival at 0.71 au at 16:30 UT on 2022 September 5. A clear velocity dispersion can be seen in Figures 3(b) and (c), showing the SPR time (about 16:14 UT) very close to the CME eruption time (A. Kouloumvakos et al. 2025). These observations suggest that SolO was well connected to the CME-driven shock at the beginning of the eruption and the SEPs underwent rapid acceleration near the Sun.

This event presents a unique opportunity to investigate the radial and longitudinal variations of widespread SEP events and raises several critical questions: (1) What physical mechanism explains the IVA observed by PSP? (2) What causes the delayed SEP onset at PSP? (3) What drives the increase in $E_{\mathrm{Max}}$ downstream at PSP? (4) Why do the time–intensity profiles and $E_{\mathrm{Max}}$ values (approximately 20 MeV at PSP versus over 100 MeV at SolO) differ so much in the upstream region, despite their relatively small longitudinal separation? To address these questions, it is essential to simulate both SEP acceleration and transport from the near-Sun region to the observers based on realistic setups.

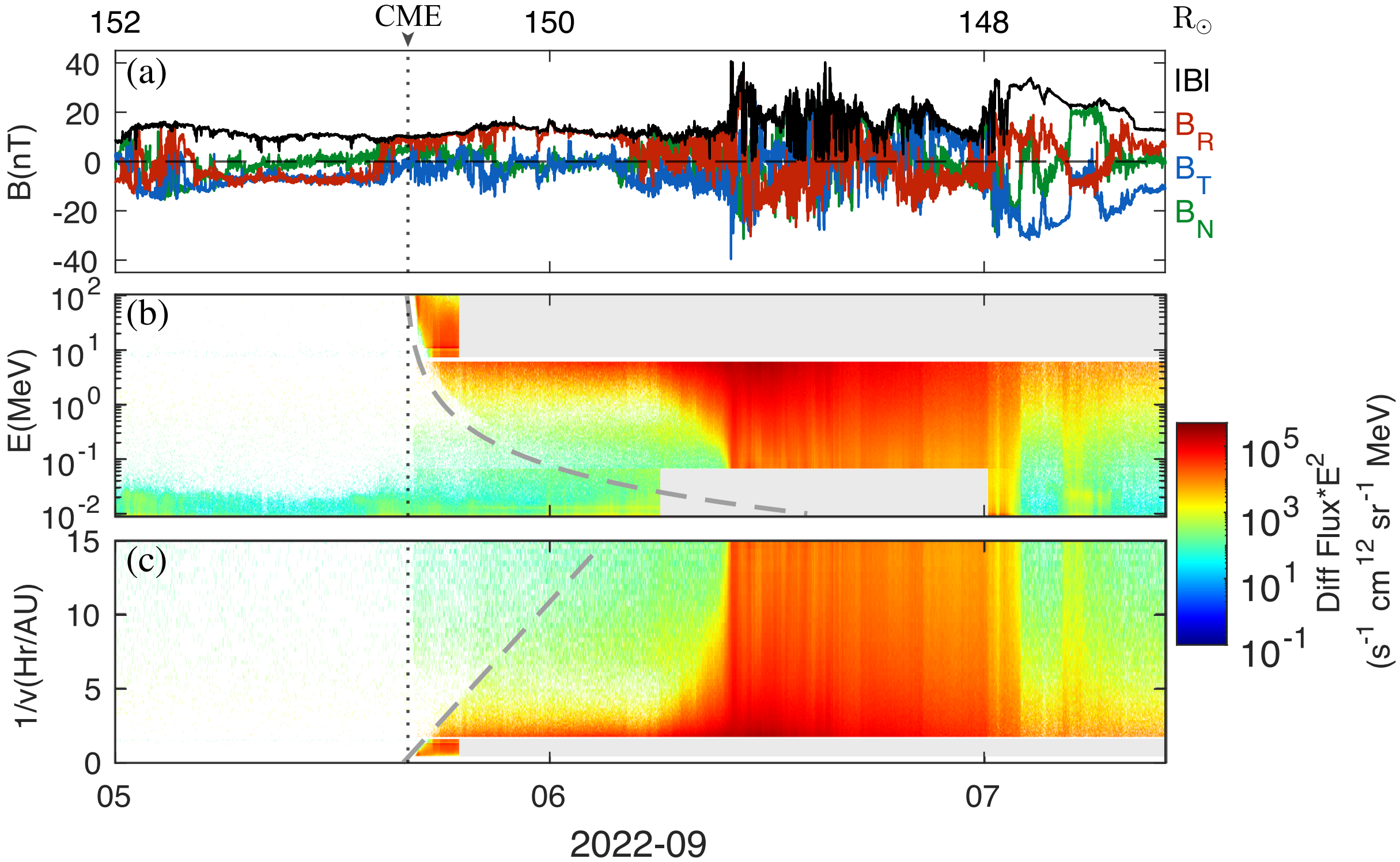

**Figure 3.** Overview of the SolO observations from 2022 September 5 00:00 UT to September 7 10:00 UT. (a) Magnetic field magnitude and components in the RTN coordinate system, with the heliocentric distance of SolO shown along the top $x$-axis in solar radii. (b) Integrated energy spectrograms for ion intensities from EPD, including sunward HET, sunward EPT, and STEP. (e) Energy spectrograms for EPD plotted against inverse velocity. The vertical dotted line indicates the CME eruption time. The gray dashed lines represent the fitted velocity dispersion.

## 3. Modeling

We employ a newly developed SEP model—PARMISAN—to simulate SEP observations for both PSP and SolO. PARMISAN is a data-driven and physics-based model designed to simulate the acceleration and transport processes of SEPs within the heliosphere. The model integrates the test particle model developed at the University of Arizona (e.g., F. Guo et al. 2010; J. Kóta 2010; J. Giacalone 2015; X. Chen et al. 2022) with the Space Weather Modeling Framework (SWMF; G. Tóth et al. 2005, 2012; T. I. Gombosi et al. 2021) developed at the University of Michigan. Specifically, the background solar wind is simulated using the Alfvén Wave Solar atmosphere Model (-Realtime) (AWSoM/AWSoM-R; I. V. Sokolov et al. 2013, 2021; B. van der Holst et al. 2014; T. I. Gombosi et al. 2018), and the CME eruptions are modeled with the Eruptive Event Generator Gibson and Low model (EEGGL; S. E. Gibson & B. C. Low 1998; W. B. Manchester et al. 2004a, 2004b; D. Borovikov et al. 2017; M. Jin et al. 2017a, 2017b) in the SWMF. Adaptive mesh refinement (AMR) is applied to improve the spatial resolution along the CME propagation direction. The shock surface is captured with the hyperbolic tangent function. Suprathermal particles are injected at shocks, and particle acceleration and transport are simulated by numerically solving the Parker transport equation via the stochastic integration method. This integrated model is designed to investigate the spatial and temporal evolution of SEPs throughout the heliosphere. The following sections provide a detailed description of the simulation setup.

### *3.1. Background Solar Wind*

AWSoM is a 3D global magnetohydrodynamic (MHD) model driven by the radial component of the photospheric magnetic field. This model provides a solution for the ambient solar wind extending from the upper solar chromosphere to the corona and heliosphere. The extended MHD equations are solved using the numerical Block Adaptive Tree Solar Wind Roe-type Upwind Scheme (BATS-R-US; K. G. Powell et al. 1999). The inner boundary of the magnetic field is specified by the solar synoptic/synchronic magnetic field maps, e.g., the Global Oscillation Network Group (GONG; J. W. Harvey et al. 1996). The initial magnetic field configuration is computed using the Potential Field Source Surface (PFSS) model (M. D. Altschuler & G. Newkirk 1969; K. H. Schatten et al. 1969), implemented via a finite difference method (G. Tóth et al. 2011). At the inner boundary, the proton and electron temperatures are uniformly set to $5 \times 10^4$ K, and the proton number density is set to $2 \times 10^{17}\ \mathrm{m}^{-3}$. AWSoM self-consistently incorporates low-frequency Alfvén turbulence, proton temperature anisotropy (accounting for parallel and perpendicular proton temperatures), heat conduction, and radiative cooling. The energy densities of the outbound Alfvén waves are imposed at the inner boundary, with the Poynting flux scaling proportional to the surface magnetic field strength. The solar wind is heated through the nonlinear dissipation of counterpropagating Alfvén waves and accelerated by thermal and Alfvén-wave pressure gradients (X. Meng et al. 2015).

The plasma evolution between the solar surface and the top of the transition region can be simplified into a 1D problem by assuming that the magnetic field is potential and varies slowly in time, called the Threaded-Field-Line Model (TFLM; T. I. Gombosi et al. 2018; I. V. Sokolov et al. 2021). AWSoM-R is a version of the AWSoM model using TFLM inner boundary conditions. The "R" indicates it can run faster than real time on ~200 cores, with a grid resolution of ~2° near the Sun. A detailed description of the AWSoM-R model and its underlying physics can be found in I. V. Sokolov et al. (2021).

AWSoM/AWSoM-R has been validated extensively by comparing the simulated solar wind properties with in situ observations (E. Wraback et al. 2024; L. Zhao et al. 2024) and remote sensing data of the solar corona in various wavelengths (N. Sachdeva et al. 2019, 2021) in different phases of the solar cycle (Z. Huang et al. 2023, 2024a), as well as comparing with magnetic field topology provided from the PFSS model (Z. Huang et al. 2024b). These validation efforts ensure that AWSoM-R provides a reliable representation of the background solar wind and large-scale heliospheric structures, such as HCSs and corotating interaction regions (CIRs), which play critical roles in the acceleration and transport of SEPs in interplanetary space.

We simulate the pre-event ambient magnetic field and solar wind plasma using the AWSoM-R model, driven by the hourly updated ADAPT-GONG magnetogram (K. S. Hickmann et al. 2015) at 16:00 UT on 2022 September 5 (Figure 4(a)). The simulation domain spans from the solar corona (1.1–20 $R_\odot$) to the inner heliosphere (20 $R_\odot$–1 au). Figure 4(b) illustrates the distribution of solar wind velocity ($U_{\rm SW}$) in the ecliptic plane ($Z = 0$) within the heliocentric inertial coordinate system. The blue, red, and green dots denote the positions of PSP, SolO, and Earth, respectively. Solid lines represent the magnetic field lines connected to each spacecraft and Earth. At this time, PSP and SolO were situated in different solar wind regimes: PSP was embedded in fast solar wind (>400 km s$^{-1}$), while SolO was located within slow solar wind (<400 km s$^{-1}$). These spatial variations in the solar wind speed directly modulate the local shock strength and thereby contribute to the distinct SEP energy spectra and intensity profiles observed by PSP and SolO.

### 3.2. CME Initiation

The CME eruption and associated solar wind disturbances are simulated using the AWSoM-R model with the parameters provided by EEGGL. The CME is triggered by inserting an unstable flux rope into the steady-state solar corona solution obtained from AWSoM-R. This setup enables the immediate eruption of the CME due to the force imbalance between the background plasma pressure and the magnetic pressure of the flux rope. The key parameters of the flux rope, including the magnetic field strength, position, and helicity, are derived from real-time GONG magnetograms and direct coronagraphic observations of CMEs (M. Jin et al. 2017a). Notably, the magnetic field strength is constrained by the observed CME speed near the Sun. To optimize this parameter, successive runs were made to give the best CME arrival time at different observers. The combined EEGGL and AWSoM-R approach facilitates a data-driven simulation of CME evolution in both the solar corona and interplanetary medium, which has been validated and studied in previous works (e.g., M. Jin et al. 2017a, 2017b).

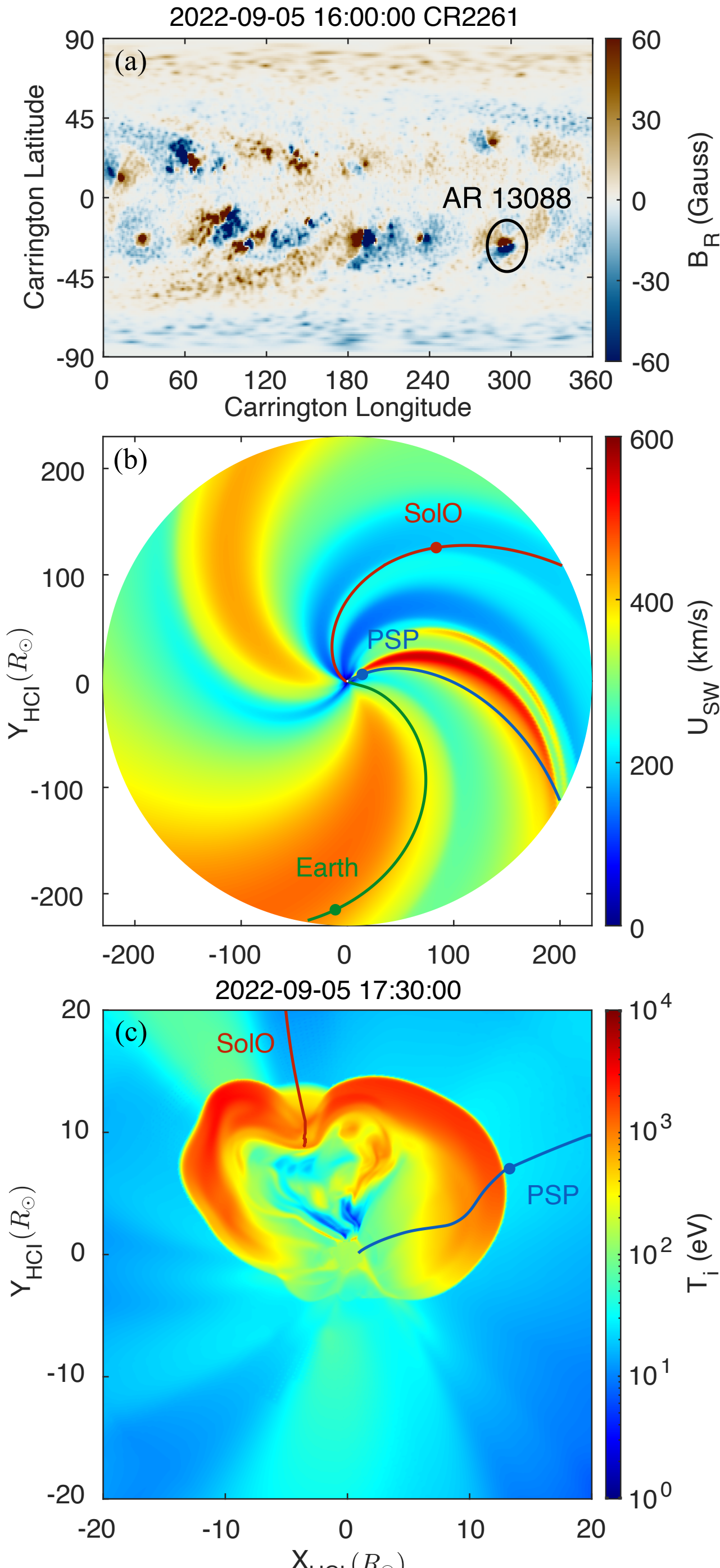

**Figure 4.** (a) Radial photospheric magnetic field $B_R$ from the ADAPT-GONG magnetogram for Carrington rotation (CR) 2261. The magnetic field saturation value is set to 60 G to enhance the visibility of active region AR 13088. (b) Steady-state solar wind total velocity in the $Z = 0$ plane of the heliocentric inertial coordinate system at 16:00 UT on 2022 September 5. (c) Configuration of the CME embedded within the background solar wind solution (shown in the ion temperature $T_i$) at 17:30 UT, when the shock front is approximately crossing the PSP. The blue, red, and green dots denote the positions of PSP, SolO, and Earth, respectively. Solid lines represent the magnetic field lines connected to each spacecraft and Earth.

Figure 4(c) presents the CME configurations in the ion temperature $T_i$ at 17:30 UT on 2022 September 5, when the CME crossed the PSP around 15 $R_\odot$. At this time, SolO was magnetically connected to the nose of the CME-driven shock, as indicated by the red line in Figure 4(c), very close to the HCS. The CME was deformed here due to the high density of solar wind plasma in this region, resulting in an oblique shock geometry. Meanwhile, PSP was initially connected to the east flank of the shock. As PSP approached its perihelion, the intersection point between the shock surface and magnetic field line connecting to PSP (referred to as the COBPOINT hereafter, short for Connecting with the OBserver POINT, following A. M. Heras et al. 1995) gradually shifted along the shock surface.

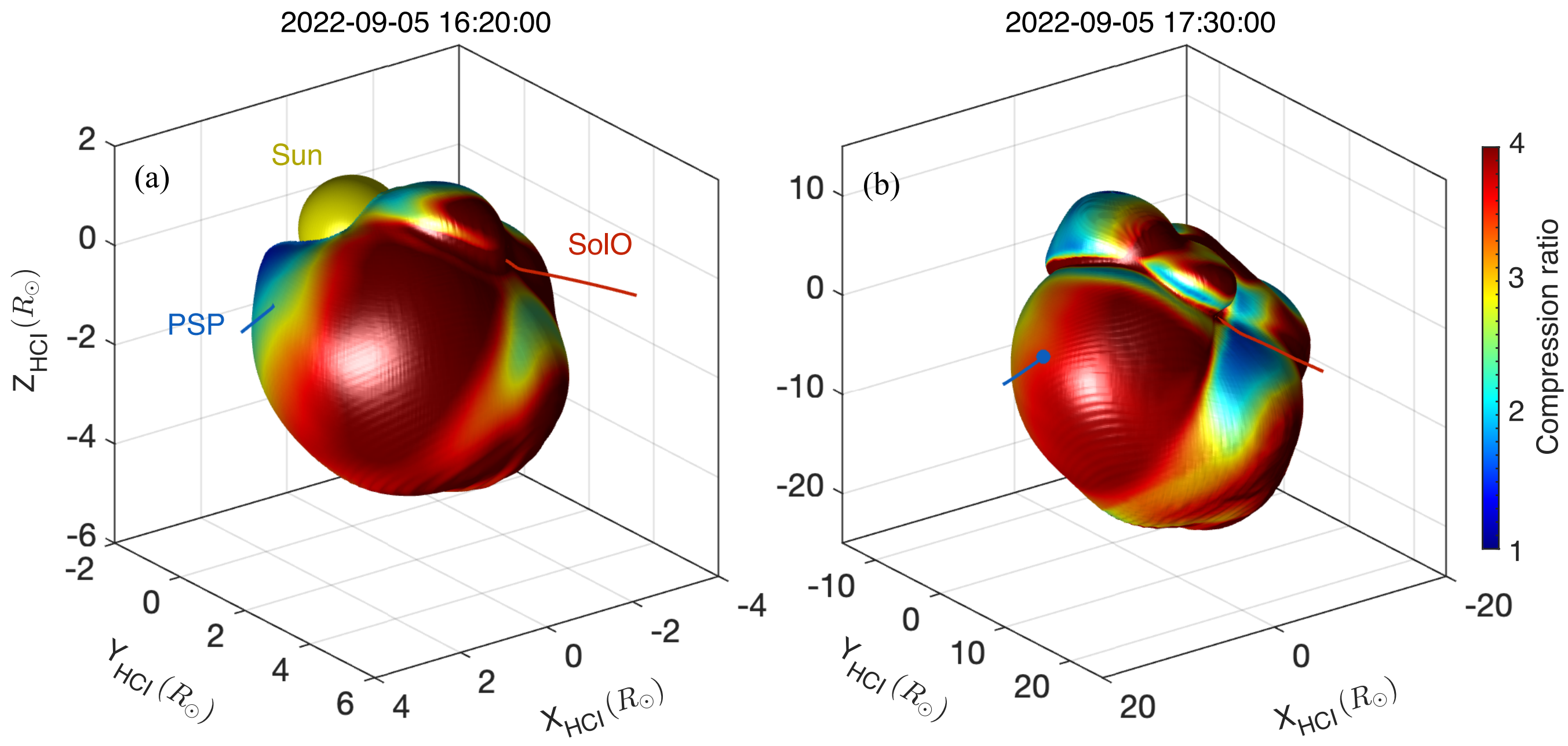

**Figure 5.** The distribution of compression ratio ($\rho_2/\rho_1$) along the shock surfaces at (a) 16:20 UT and (b) 17:30 UT on 2022 September 5.

### 3.3. Shock Capture

Accurate modeling of SEP acceleration requires the shock layer to be sufficiently resolved, with the shock thickness $\delta$ being much smaller than the diffusive skin depth of energetic particles. Achieving such a high spatial resolution remains a significant challenge in MHD simulations of large SEP events. To overcome this limitation, we employ the shock-capturing and sharpening tools to reconstruct a refined, thin shock profile, ensuring that the shock layer is resolved with the necessary precision (see details in Section 4.3 of W. Liu et al. 2025). First, we implement AMR to enhance spatial resolution in the vicinity of CMEs, which dynamically adjusts the grid resolution based on predefined criteria. AMR allows for the refinement or coarsening of the grid by one or more levels, depending on the specific resolution requirements. Several physics-based criteria, including pressure gradients, velocity divergence, and relative density changes, are used to determine where and how AMR is applied. The shock front is identified by locating the point of the maximum plasma velocity divergence (multiplied by cell size) along the radial direction starting from the longitude–latitude grid. Once the shock front is captured, the shock normal direction is estimated at each grid point along the shock front based on the adjacent grid cells. The shock thickness is refined locally by fitting the upstream and downstream plasma properties of the MHD shock to a hyperbolic function, characterized by a length scale $\delta_{sh}$. In the local shock frame, the fluid velocity along the shock normal can be approximated by a hyperbolic tangent function,

$$U(x') = \frac{U_1 + U_2}{2} - \frac{U_1 - U_2}{2}\tanh\left(\frac{x'}{\delta_{sh}}\right), \tag{2}$$

where $U_1 = V_{sh}$ and $U_2 = V_{sh}/s$. $V_{sh}$ is the shock speed, $s$ is the density compression ratio, $\delta_{sh}$ is the shock width, and $x'$ is the distance to the shock front along the shock normal. Figure 5 presents the shock surface captured 10 and 80 minutes after the CME eruption on 2022 September 5. The deformation of the shock geometries is primarily modulated by the upstream solar wind density, particularly in regions influenced by HCSs or CIRs. The distribution of the compression ratio along the shock surface indicates that the acceleration rate can vary significantly across different regions of the shock front, in association with large-scale heliospheric structures. These variations contribute to an initial spatial diversity in the source populations of SEPs, which subsequently leads to noticeable spatial variations in SEP fluxes near 1 au.

### 3.4. Particle Acceleration and Transport

We simulate SEP acceleration and transport by numerically solving the Parker transport equation (E. N. Parker 1965). This equation incorporates most macroscopic transport processes, including spatial diffusion in turbulent magnetic fields, advection with the solar wind, magnetic gradient or curvature drifts, and adiabatic energy changes due to plasma compression or rarefaction. In this work, we consider the 1D Parker equation of a spherically symmetric geometry,

$$\frac{\partial f}{\partial t} = \frac{1}{r^2}\frac{\partial}{\partial r}\left(r^2\kappa_{rr}\frac{\partial f}{\partial r}\right) - U\frac{\partial f}{\partial r} + \frac{1}{3r^2}\frac{\partial(r^2U)}{\partial r}\frac{\partial f}{\partial \ln p} + Q, \tag{3}$$

where $f$ is the phase-space distribution function, $r$ is the heliocentric distance, $\kappa_{rr}$ is the diffusion coefficient in the radial direction, $U$ is the solar wind speed, and $Q$ is the source term. Equation (3) assumes SEPs of quasi-isotropic pitch-angle distributions in the plasma frame, which is always achieved for the SEPs near the shock where the charged particles have undergone sufficient diffusion. While particle acceleration at planar shocks depends only on the normal components of plasma parameters (Equation (8)), our spherical shock model incorporate geometric effects by resolving these shock normal components at each time step. This approach remains valid when the radius of the shock curvature is much larger than the particle diffusive skin depth. By updating these resolved parameters as the shock propagates, shock geometric evolution in SEP acceleration processes is considered in our simulation. The plasma flow speed in the local shock frame depends on the shock normal angle $\theta_{\rm Bn}$: $U_1 = V_{\rm sh} - \cos\theta_{\rm Bn} V_{\rm SW}$ and $U_2 = U_1/s$. $\kappa_{\rm rr}$ will also evolve with $\theta_{\rm Bn}$ as the shock expands in the corona (J. R. Jokipii & E. N. Parker 1970):

$$\kappa_{rr} = \kappa_{\perp} + (\kappa_{\parallel} - \kappa_{\perp}) \cos^2\theta_{\rm Bn}. \quad (4)$$

The parallel diffusion coefficient $\kappa_{\parallel}$ follows the empirical formula obtained from PSP observations (X. Chen et al. 2024),

$$\kappa_{\parallel} = \kappa_0\, r^{1.17}\, E^{0.71}, \quad (5)$$

where the heliocentric distance $r$ and the particle's energy $E$ are in units of au and keV. $\kappa_0$ is chosen here as $10^{18}$ cm$^2$ s$^{-1}$, and $\kappa_{\perp}$ is assumed to be 3% of $\kappa_{\parallel}$. We choose protons as the source particles with sufficiently high energy ($E_0 =$ 50 keV) so that the pitch-angle distribution is almost isotropic and the shock geometry has less impact on the injection rate.

Equation (3) can be written in the form of a Fokker–Planck equation and solved by successively integrating trajectories of pseudoparticles with the corresponding stochastic differential equations (e.g., C. Pei et al. 2010),

$$\Delta r = r_1\sqrt{2\kappa_{rr}\Delta t} + U\Delta t + \frac{1}{r^2}\frac{\partial(r^2\kappa_{rr})}{\partial r}\Delta t, \quad (6)$$

$$\Delta p = -\frac{p}{3r^2}\frac{\partial(r^2 U)}{\partial r}\Delta t, \quad (7)$$

where $R_1$ is the normalized random numbers satisfying $\langle r_1\rangle = 0$ and $\langle r_1^2\rangle = 1$. The pseudoparticles of 50 keV are continuously injected to the shock front, and the intensity falls off with the radial distance as $Q \propto r^{-2}$. To ensure the acceleration is simulated properly, we consider a sharp layer of shock ($\delta_{\rm sh}$ in Equation (2)) that is much smaller than the characteristic length of diffusion acceleration $\kappa_{\rm rr}/U_1$.

The magnetic field and plasma properties in the upstream and downstream regions are obtained from the AWSoM-R model along the field lines connecting to each spacecraft. These plasma parameters in the shock layer are reconstructed by the shock-capturing and sharpening tools introduced above. The simulation starts when the shock establishes the magnetic connection to the spacecraft and ends when the CME is beyond 1 au. The boundaries are the same as those in the AWSoM-R model, and the time step is small enough to resolve the motion of pseudoparticles in the shock layer. The particles will be removed if they enter the inner or outer boundaries. We also use the particle splitting technique (J. Giacalone 2005) to improve the statistics for high-energy particles.

## 4. Results

The 3D shock surfaces are captured with a 1 minute cadence since the eruption. Plasma parameters along the shock front are derived from the time-accurate AWSoM-R simulations, while shock velocities are determined by tracking the displacements between successive shock surfaces. Figure 6 displays the temporal evolution of shock and plasma parameters at the PSP (blue) and SolO (red) COBPOINTs as the shock expands outward from the Sun. Panels show, from top to bottom, (a) upstream solar wind velocity, (b) shock speed, (c) Alfvén Mach number, (d) shock normal angle, and (e) density compression ratio. Substantial differences in these parameters are observed between PSP and SolO. The COBPOINT of SolO lies in the slow solar wind near the HCS, where small magnetic field strength, reduced solar wind speeds, and large plasma density collectively enhance the Alfvén Mach number and shock compression ratio locally. These conditions facilitate rapid SEP acceleration locally, producing the enhanced flux energies and rapid onset as observed by SolO in Figure 3.

In contrast, Figure 5(a) reveals that PSP was initially connected to the shock flank, characterized by higher upstream solar wind speeds in Figure 6(a). This configuration suppresses the shock strength (compression ratio), leading to smaller SEP acceleration rates compared to SolO's observations, consistent with what was shown in A. Kouloumvakos et al. (2025). Consequently, it requires more time for SEPs to be accelerated to the higher energies observed by PSP. The prolonged acceleration timescale under these conditions gives rise to the IVA observed in Figure 2(d), where lower-energy particles arrive earlier than their higher-energy counterparts. Note that these IVA features have also been reproduced in previous numerical models that are consistent with our results (e.g., G. P. Zank et al. 2000; G. Li et al. 2003; W. K. M. Rice et al. 2003). After shock passage, PSP's high perihelion velocity allowed the magnetic connectivity to quickly shift to a region of enhanced shock compression. This transition permitted the higher-energy SEPs accelerated in the stronger compression zone to propagate toward the Sun along open magnetic field lines. Consistent with this scenario, PSP detected strongly sunward-dominated SEP fluxes during this energy-rise phase (C. M. S. Cohen et al. 2024). Collectively, these dynamics explain the unusual downstream energy increase observed in Figure 2(b).

To understand the underlying mechanisms that generate these features, we apply the PARMISAN model to simulate the SEP observations at PSP and SolO. We update the background solar wind and shock properties every minute along the COBPOINT of each spacecraft. Figure 7(a) displays the simulated spectrogram of proton intensities observed near the PSP orbit at 15 $R_\odot$. Figure 7(b) compares the simulated time–intensity profiles at PSP's orbit with EPI-Lo observations across selected energy channels. Note that we employ the empirical diffusion coefficient model (Equation (5)) without optimizing $\kappa$ to achieve a perfect match with observations. Event-by-event parameter optimization would be required for an exact fit, which is beyond the scope of this paper. Unlike typical SEP events, the spectrogram exhibits a distinct rounded-nose feature near 2 MeV. Below this threshold, particle arrival follows standard velocity dispersion (higher-energy particles arriving first followed by lower-energy ones), while above this energy, the trend inverts: lower-energy particles precede the higher-energy counterparts. This inversion arises because particles require more time to reach higher energies when acceleration rates are small. Considering the space- and time-dependent DSA, the acceleration rate at the shock in a spherical geometry is given by (J. R. Jokipii 1987; L. O. Drury 1991)

$$\frac{1}{p}\frac{dp}{dt} = \frac{U_1(R_{\rm sh}) - U_2(R_{\rm sh})}{3} \times \left[\int_{R_{\rm min}}^{R_{\rm max}} dr' \exp\left(-\left|\int_{R_{\rm sh}}^{r'} \frac{U(r)}{\kappa(r, p)} dr\right|\right)\right]^{-1}. \tag{8}$$

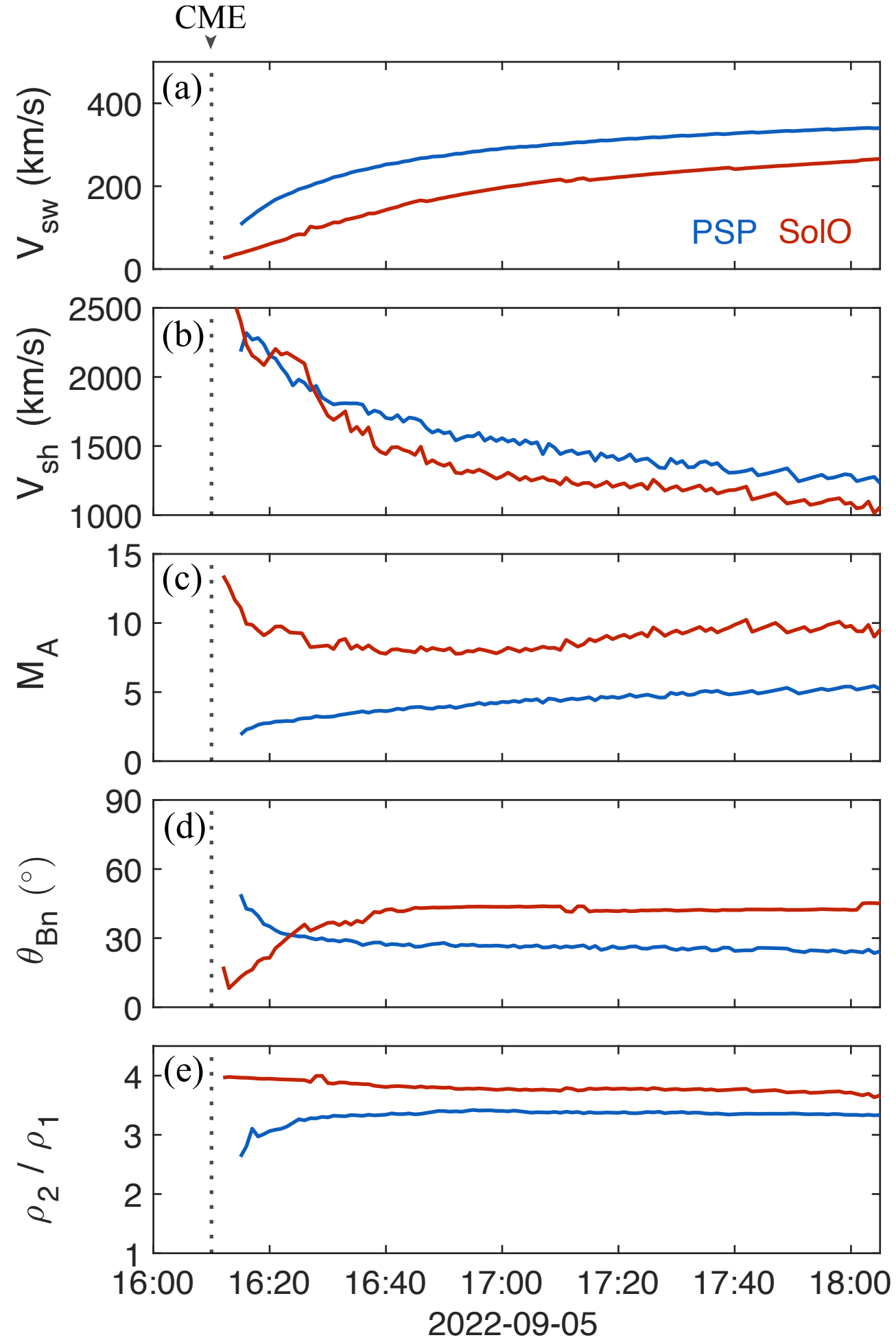

**Figure 6.** (a) Solar wind velocity, (b) shock speed, (c) Alfvén Mach number, (d) shock normal angle, and (e) density compression ratio estimated at the COBPOINTs of PSP (blue) and SolO (red) as the shock expands outward from the Sun. The vertical dotted line marks the eruption time of the CME.

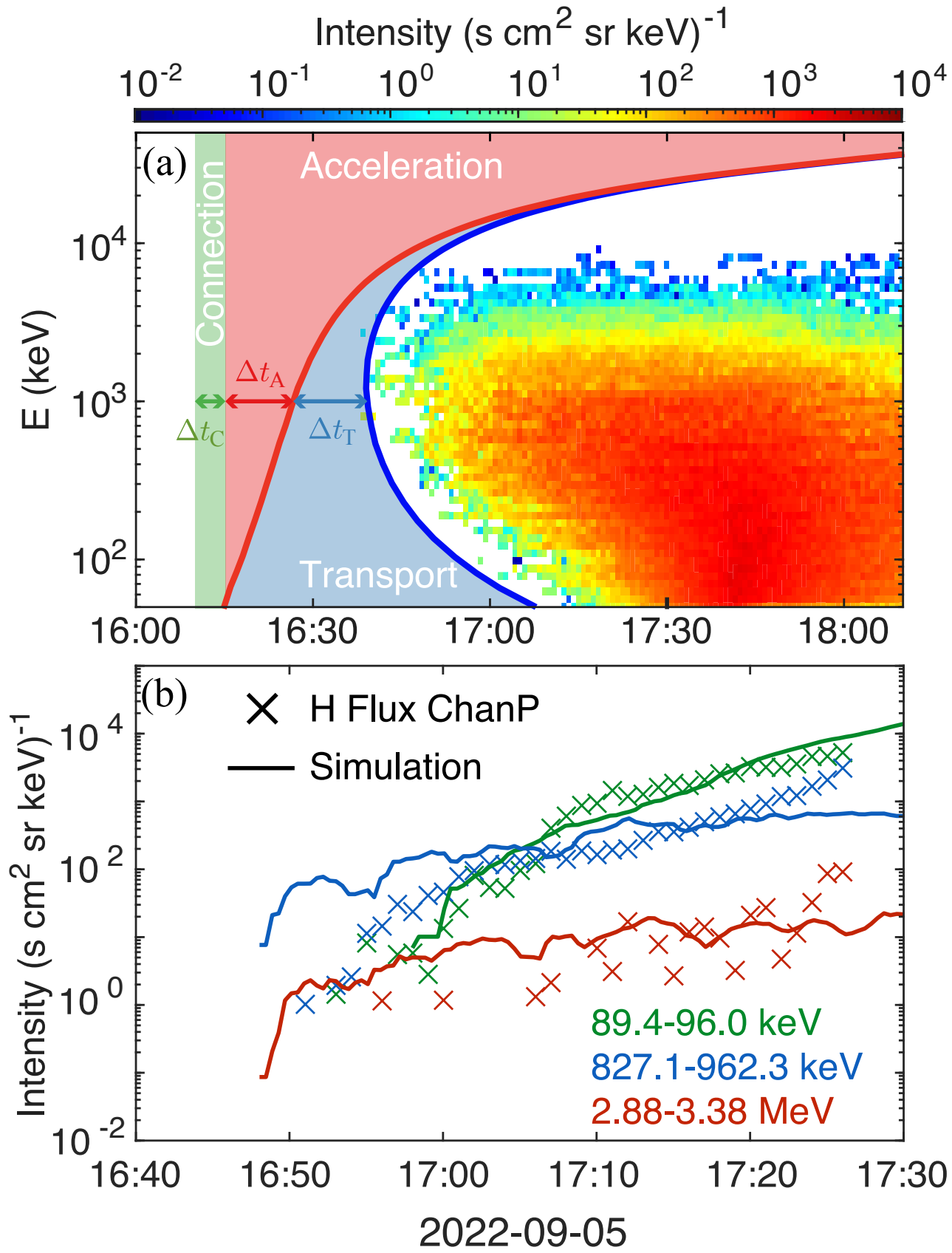

**Figure 7.** (a) Simulated spectrogram of proton intensities observed near the PSP orbit at 15 $R_\odot$. The green shaded region marks the interval between the CME eruption and the time PSP first becomes magnetically connected to the shock. The red and blue shaded regions indicate the durations of proton acceleration at the shock and transport to the observer, respectively. The red curve tracks the maximum proton energy achievable at the shock as it expands outward. The blue curve outlines the spectrogram limits imposed by the magnetic connection, acceleration, and transport processes. Double-headed arrows indicate the respective timescales ($\Delta t_C$, $\Delta t_A$, and $\Delta t_T$) for 1 MeV protons. (b) Comparison of simulated (solid lines) time–intensity profiles and PSP's observations (cross symbols) for a selection of energy channels.

Similar to Equation (1), $U_1(R_{\rm sh})$ and $U_2(R_{\rm sh})$ are the upstream and downstream flow speeds estimated at the shock. $R_{\rm min}$ and $R_{\rm max}$ are the same as the inner and outer boundaries of the AWSoM-R model. To estimate the maximum momentum $p_{\rm max}$ attainable from an initial momentum $p_0$ over a time interval $t_0$ to $t$, we integrate

$$p_{\rm max}(t) = p_0 + \int_{t_0}^{t} \frac{U_1(R_{\rm sh}) - U_2(R_{\rm sh})}{3} \times \left[\int_{R_{\rm min}}^{R_{\rm max}} dr' \exp\left(-\left|\int_{R_{\rm sh}}^{r'} \frac{U(r)}{\kappa(r, p)} dr\right|\right)\right]^{-1} p(t) dt, \tag{9}$$

where $t_0$ is the time when the observer first becomes magnetically connected to the shock and $t$ is any subsequent time of interest. Both plasma and shock velocities are obtained from the MHD simulation. Here, we consider the spatial dependence of $\kappa(r, p)$ near the shock and in the downstream region, following an empirical formula obtained from PSP observations (Equation (5)). In the far upstream, we increase $\kappa(r, p)$ to a value 10 times larger than in the downstream, following J. Giacalone et al. (2023). In Figure 7(a), the green shaded region indicates the interval between the eruption of the event and the time when PSP first established a magnetic connection to the shock front. The red curve shows the maximum energy achievable at the shock, as calculated from Equation (9). The red shaded region shows the time required for a particle to be accelerated from the initial energy (50 keV) to this evolving maximum energy. In this scenario, particles undergo continuous acceleration at the shock as it expands outward from the Sun. The accelerated particles are also free to escape the shock and propagate to observers at different locations. This transport duration, represented by the blue shaded region, is estimated by $\Delta t_T = (R_{\rm obs} - R_{\rm sh})/v_p$, where $R_{\mathrm{obs}}$ and $R_{\mathrm{sh}}$ denote the locations of the observers and the shock at the time the particle escapes and $v_p$ is the particle speed upon its escape. The blue curve outlines the spectrogram limits imposed by the magnetic connection, acceleration, and transport processes.

The total observed time delay $\Delta t$ for an SEP event comprises three components:

$$\Delta t = \Delta t_C + \Delta t_A + \Delta t_T. \tag{10}$$

(1) Connection time $\Delta t_C$ is the time for the observer to first establish magnetic connection to the shock front since the eruption. (2) Acceleration time $\Delta t_A$ is the time required for particles to be accelerated to a specific energy from the source energy via DSA. (3) Transport time $\Delta t_T$ is the propagation time from the shock to the observer along magnetic field lines. For example, the double-headed arrows in Figure 7(a) present the respective timescales for 1 MeV protons, where $\Delta t_A$ are comparable to $\Delta t_T$. The nose of the spectrogram corresponds to the earliest arriving particles (minimum time delay $\Delta t$). At energies below this nose, the transport process dominates $\Delta t$, producing a "regular velocity dispersion," although the acceleration process still contributes to the overall profile. At higher energies, the acceleration process dominates, resulting in an IVA feature. The nose energy increases with the observer's radial distance due to the longer transport time.

In large SEP events, strong shocks minimize $\Delta t_A$, making $\Delta t_T$ dominant for well-connected observers. However, for weak shocks or observers linked to weaker shock regions (e.g., PSP before perihelion), $\Delta t_A$ could be comparable to $\Delta t_T$, especially when the observer is close to the shock. This interplay produces the observed IVA. We note that this feature may persist even at 1 au when the observer is initially connected to a weak shock region. However, once a stable connection to the strong shock region is established, the feature becomes unlikely to appear at large heliocentric distances, as $\Delta t_T$ exceeds $\Delta t_A$ by orders of magnitude. But even in this case, the time-dependent acceleration effect can still play an important role in determining the SPR time and source localization.

Figures 8(a) and (b) present the observed and simulated energy spectrograms for SolO at 0.71 au, respectively. The maximum energy and onset time show a close match with the first-arriving particles. To achieve this consistency, we reduced $\kappa_0$ to one-third of the value used for PSP, likely due to the enhanced turbulence condition in the upstream at SolO (D. Trotta et al. 2024). Note that the diffusion coefficient distribution, derived from average solar wind conditions, can vary locally by an order of magnitude (X. Chen et al. 2024). The discrepancy for the later-arriving particles occurs because these particles were accelerated earlier and have propagated back to the observer from a large distance. The finite scale of our simulation domain does not encompass the full transport path of these particles; therefore, the modeled maximum energy for this population is naturally lower than observed. Additionally, particle drifts, plasma instabilities, and magnetic reconnection during shock–current sheet interactions (e.g., with the HCS) can trap SEPs locally, enabling further acceleration (J. Kota & J. R. Jokipii 1994; J. Giacalone & D. Burgess 2010; B. Cerutti & G. Giacinti 2020; M. Nakanotani et al. 2021).

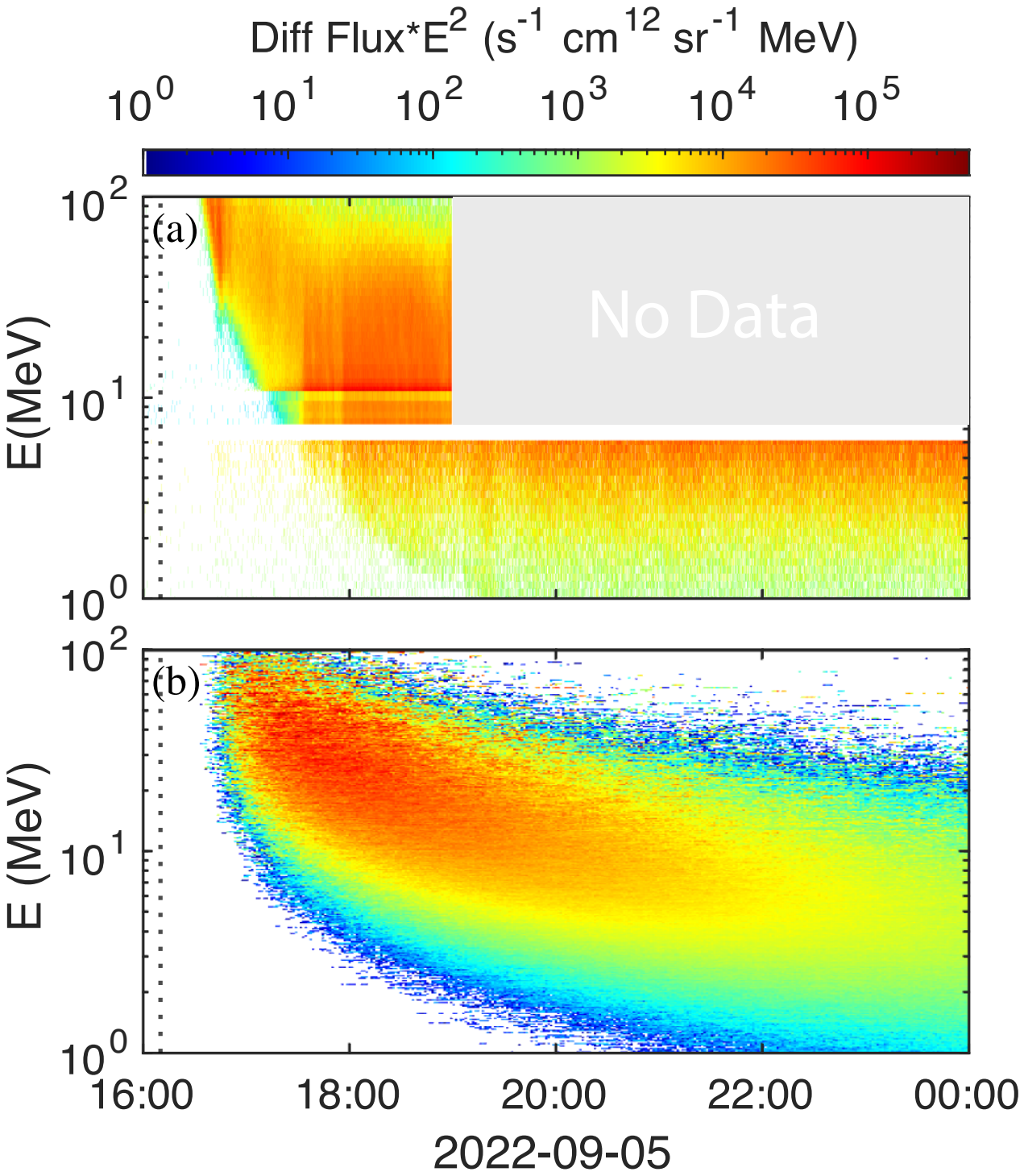

**Figure 8.** Modeled SEP intensities at SolO. (a) Integrated energy spectrograms for ion intensities from EPD. (b) The simulated spectrograms at the SolO location. Vertical dashed lines denote the eruption time in both panels.

## 5. Conclusions

This study investigates PSP and SolO observations of the 2022 September 5 SEP event with physics-based simulations using the PARMISAN model. Our simulations reveal that PSP was initially magnetically connected to the shock flank—a region of lower compression ratio and weaker shock strength—where slower acceleration rates delayed the energization of SEPs to energies >1 MeV. The acceleration timescale became comparable to the transport timescale, producing the observed IVA as high-energy particles arrived later than lower-energy ones. This mechanism also explains the delayed SEP onset at PSP. As PSP traversed its high-speed trajectory near perihelion, its magnetic connectivity shifted dynamically postshock to regions of stronger compression, significantly enhancing subsequent particle energization. A subset of these high-energy particles streamed backward along the IMF toward the Sun, detected by PSP with a pronounced sunward anisotropy (C. M. S. Cohen et al. 2024). This process provides a potential explanation for the downstream energy increase in PSP observations. In contrast, SolO was magnetically linked to a strong shock region in the slow solar wind, providing access to efficient acceleration of high-energy particles due to favorable shock conditions. The primary results are summarized below.

1. *IVA at PSP*. The delayed arrival of higher-energy (>1 MeV) particles relative to lower-energy populations (Figure 2) arises from time-dependent DSA. The initial low compression ratios and slow acceleration rates at the COBPOINT (Figure 6) of PSP extended the acceleration time $\Delta t_A$, which is comparable to $\Delta t_T$.

2. *Radial and longitudinal SEP variations*.
   - SolO's connection to the nose of the shock provided optimal conditions for efficient accelerations of SEPs. These factors enabled detection of $>$100 MeV particles at SolO shortly after the CME eruption (Figure 3). These particles show a regular velocity dispersion without an obvious delay of the particle release time.
   - PSP's initial connection to the weak shock region at the east flank resulted in slower acceleration and relatively lower-energy SEPs ($\sim$20 MeV). However, its downstream connectivity shifted to a strong compressed shock region after perihelion characterized by the sunward-propagating high-energy particles, which leads to the postshock $E_{\mathrm{Max}}$ increase (Figure 2).
3. *Delayed onset at PSP*. The SEP onset delay at PSP ($\sim$35 minutes posteruption) decomposes into $\Delta t = \Delta t_C + \Delta t_A + \Delta t_T$. $\Delta t_A$ contributed considerably to the total delay as the initial acceleration rate was small.

These results demonstrate that SEP variability near the Sun is governed by the local shock evolution, originating from dynamic interactions between the CME-driven shock and upstream solar wind conditions. These differences modulate local acceleration rates across the shock surface, thereby driving the spatial disparities in SEP energies and intensities. Also, we find that these shock and solar wind parameters evolve in time as the shock expands outward from the Sun. This evolution emphasizes that the acceleration process is highly time- and space-dependent, especially when the acceleration timescale is comparable to the transport timescale.

## Acknowledgments

This work was supported in part by the NASA LWS Strategic Capabilities project at the University of Michigan under NASA grant 80NSSC22K0892 (SCEPTER) and NASA SWxC grant 80NSSC23M0191 (CLEAR). L.Z. acknowledges NASA LWS grant 80NSSC21K0417, NASA R2O2R grant 80NSSC22K0269, NASA HSR grant 80NSSC23K0091, and NSF ANSWERS grant GEO-2149771. N.S. acknowledges NASA LWS grant 80NSSC24K1104, NSF Solar Terrestrial grant 2323303, and NASA R2O2R grant 80NSSC23K0450. A.K. acknowledges financial support from NASA's NNN06AA01C (Parker Solar Probe EPI-Lo) contract and NASA LWS grant 80NSSC25K0130. Parker Solar Probe was designed and built and is now operated by the Johns Hopkins Applied Physics Laboratory as part of NASA's Living with a Star (LWS) program (contract NNN06AA01C). Support from the LWS management and technical team has played a critical role in the success of the Parker Solar Probe mission. We acknowledge the PSP/IS⊙IS team (PI: David McComas, Princeton University) and the PSP/FIELDS team (PI: Stuart D. Bale, UC Berkeley) for use of data. We acknowledge the EPD and magnetic field data from Solar Orbiter, generated and maintained by the EPD team and magnetometer team on the Solar Orbiter Archive (SOAR). This work utilizes data produced collaboratively between AFRL/ADAPT and NSO/NISP. We thank the referee for helpful comments. Resources supporting this work were provided in part by the NASA High-End Computing (HEC) Program through the NASA Advanced Supercomputing (NAS) Division at Ames Research Center. The authors acknowledge the Texas Advanced Computing Center (TACC) at the University of Texas at Austin for providing HPC resources that have contributed to the research results reported in this paper.

## ORCID iDs

Xiaohang Chen https://orcid.org/0000-0003-2865-1772
Lulu Zhao https://orcid.org/0000-0003-3936-5288
Joe Giacalone https://orcid.org/0000-0002-0850-4233
Nishtha Sachdeva https://orcid.org/0000-0001-9114-6133
Igor V. Sokolov https://orcid.org/0000-0002-6118-0469
Gábor Tóth https://orcid.org/0000-0001-8459-2100
Christina M.S. Cohen https://orcid.org/0000-0002-0978-8127
David Lario https://orcid.org/0000-0002-3176-8704
Fan Guo https://orcid.org/0000-0003-4315-3755
Athanasios Kouloumvakos https://orcid.org/0000-0001-6589-4509
Tamas I. Gombosi https://orcid.org/0000-0001-9360-4951
Zhenguang Huang https://orcid.org/0000-0003-1674-0647
Ward B. Manchester, IV https://orcid.org/0000-0003-0472-9408
Bart van der Holst https://orcid.org/0000-0001-5260-3944
Weihao Liu https://orcid.org/0000-0002-2873-5688
David J. McComas https://orcid.org/0000-0001-6160-1158
Matthew E. Hill https://orcid.org/0000-0002-5674-4936
George C. Ho https://orcid.org/0000-0003-1093-2066

## References

Altschuler, M. D., & Newkirk, G. 1969, SoPh, 9, 131
Anagnostopoulos, G. C., Sarris, E. T., & Krimigis, S. M. 1986, JGRA, 91, 3020
Axford, W. I., Leer, E., & Skadron, G. 1977, ICRC, 11, 132
Bale, S. D., Goetz, K., Harvey, P. R., et al. 2016, SSRv, 204, 49
Bell, A. R. 1978, MNRAS, 182, 147
Blandford, R. D., & Ostriker, J. P. 1978, ApJL, 221, L29
Blasi, P. 2002, APh, 16, 429
Borovikov, D., Sokolov, I. V., Manchester, W. B., Jin, M., & Gombosi, T. I. 2017, JGRA, 122, 7979
Bryant, D. A., Cline, T. L., Desai, U. D., & McDonald, F. B. 1962, JGR, 67, 4983
Cane, H. V., McGuire, R. E., & von Rosenvinge, T. T. 1986, ApJ, 301, 448
Cerutti, B., & Giacinti, G. 2020, A&A, 642, A123
Chandra, R., Gopalswamy, N., Mäkelä, P., et al. 2013, AdSpR, 52, 2102
Chen, X., Giacalone, J., & Guo, F. 2022, ApJ, 941, 23
Chen, X., Giacalone, J., Guo, F., & Klein, K. G. 2024, ApJ, 965, 61
Chhiber, R., Matthaeus, W. H., Cohen, C. M. S., et al. 2021, A&A, 650, A26
Cliver, E. W., & D'Huys, E. 2018, ApJ, 864, 48
Cliver, E. W., Kahler, S. W., & Reames, D. V. 2004, ApJ, 605, 902
Cohen, C. M. S., Leske, R. A., Christian, E. R., et al. 2024, ApJ, 966, 148
Desai, M., & Giacalone, J. 2016, LRSP, 13, 3
Ding, Z., Li, G., Mason, G., et al. 2024, A&A, 681, A92
Ding, Z., Wimmer-Schweingruber, R., Kollhoff, A., et al. 2025, A&A, 696, A199
Do, T. M., Fraschetti, F., Kota, J., et al. 2025, ApJ, 979, 50
Drury, L. O. 1983, RPPh, 46, 973
Drury, L. O. 1991, MNRAS, 251, 340
Forman, M. A., & Drury, L. O. 1983, ICRC, 2, 267
Fox, N. J., Velli, M. C., Bale, S. D., et al. 2016, SSRv, 204, 7
Giacalone, J. 2005, ApJ, 624, 765
Giacalone, J. 2012, ApJ, 761, 28
Giacalone, J. 2015, ApJ, 799, 80
Giacalone, J. 2017, ApJ, 848, 123
Giacalone, J., & Burgess, D. 2010, GeoRL, 37, L19104
Giacalone, J., Cohen, C. M. S., McComas, D. J., et al. 2023, ApJ, 958, 144
Gibson, S. E., & Low, B. C. 1998, ApJ, 493, 460
Gieseler, J., Dresing, N., Palmroos, C., et al. 2023, FrASS, 9, 384
Gombosi, T. I., Chen, Y., Glocer, A., et al. 2021, JSWSC, 11, 42
Gombosi, T. I., van der Holst, B., Manchester, W. B., & Sokolov, I. V. 2018, LRSP, 15, 4
Guo, F., Jokipii, J. R., & Kota, J. 2010, ApJ, 725, 128
Harvey, J. W., Hill, F., Hubbard, R. P., et al. 1996, Sci, 272, 1284
Heras, A. M., Sanahuja, B., Lario, D., et al. 1995, ApJ, 445, 497

Hickmann, K. S., Godinez, H. C., Henney, C. J., & Arge, C. N. 2015, SoPh, 290, 1105
Hill, M. E., Mitchell, D. G., Andrews, G. B., et al. 2017, JGRA, 122, 1513
Horbury, T. S., O'Brien, H., Carrasco Blazquez, I., et al. 2020, A&A, 642, A9
Howard, R. A., Moses, J. D., Vourlidas, A., et al. 2008, SSRv, 136, 67
Huang, Z., Tóth, G., Huang, J., et al. 2024a, ApJL, 965, L1
Huang, Z., Tóth, G., Sachdeva, N., & van der Holst, B. 2024b, ApJ, 965, 1
Huang, Z., Tóth, G., Sachdeva, N., et al. 2023, ApJL, 946, L47
Jin, M., Manchester, W. B., van der Holst, B., et al. 2017a, ApJ, 834, 172
Jin, M., Manchester, W. B., van der Holst, B., et al. 2017b, ApJ, 834, 173
Jokipii, J. R. 1982, ApJ, 255, 716
Jokipii, J. R. 1987, ApJ, 313, 842
Jokipii, J. R., & Parker, E. N. 1970, ApJ, 160, 735
Kahler, S. 1994, ApJ, 428, 837
Kahler, S. W., Sheeley, N. R., Jr., Howard, R. A., et al. 1984, JGRA, 89, 9683
Kallenrode, M. B., Wibberenz, G., Kunow, H., et al. 1993, SoPh, 147, 377
Kong, X., Guo, F., Giacalone, J., Li, H., & Chen, Y. 2017, ApJ, 851, 38
Kóta, J. 2010, ApJ, 723, 393
Kota, J., & Jokipii, J. R. 1994, ApJ, 429, 385
Kouloumvakos, A., Rouillard, A. P., Wu, Y., et al. 2019, ApJ, 876, 80
Kouloumvakos, A., Wijsen, N., Jebaraj, I. C., et al. 2025, ApJ, 979, 100
Krucker, S., Hurford, G. J., Grimm, O., et al. 2020, A&A, 642, A15
Krymskii, G. F. 1977, DoSSR, 234, 1306
Laitinen, T., Huttunen-Heikinmaa, K., Valtonen, E., & Dalla, S. 2015, ApJ, 806, 114
Lario, D., Kwon, R. Y., Riley, P., & Raouafi, N. E. 2017, ApJ, 847, 103
Lario, D., Kwon, R. Y., Vourlidas, A., et al. 2016, ApJ, 819, 72
Li, G., Zank, G. P., & Rice, W. K. M. 2003, JGRA, 108, 1082
Li, Y., Guo, J., Pacheco, D., et al. 2025, NSRev, 12, 10
Liu, W., Kong, X., Guo, F., et al. 2023, ApJ, 954, 203
Liu, W., Sokolov, I. V., Zhao, L., et al. 2025, ApJ, 985, 82
Long, D. M., Green, L. M., Pecora, F., et al. 2023, ApJ, 955, 152
Manchester, W. B., Gombosi, T. I., Roussev, I., et al. 2004a, JGRA, 109, A01102
Manchester, W. B., Gombosi, T. I., Roussev, I., et al. 2004b, JGRA, 109, A02107
McComas, D. J., Alexander, N., Angold, N., et al. 2016, SSRv, 204, 187
McKenzie, J. F., & Voelk, H. J. 1982, A&A, 116, 191
Meng, X., van der Holst, B., Tóth, G., & Gombosi, T. I. 2015, MNRAS, 454, 3697
Mitchell, J. G., de Nolfo, G. A., Christian, E. R., et al. 2024, ApJ, 968, 33
Müller, D., St. Cyr, O. C., Zouganelis, I., et al. 2020, A&A, 642, A1
Nakanotani, M., Zank, G. P., & Zhao, L.-L. 2021, ApJ, 922, 219
Paouris, E., Vourlidas, A., Kouloumvakos, A., et al. 2023, ApJ, 956, 58
Parker, E. N. 1965, P&SS, 13, 9
Patel, R., West, M. J., Seaton, D. B., et al. 2023, ApJL, 955, L1
Pei, C., Bieber, J. W., Burger, R. A., & Clem, J. 2010, JGRA, 115, A12107
Powell, K. G., Roe, P. L., Linde, T. J., Gombosi, T. I., & De Zeeuw, D. L. 1999, JCoPh, 154, 284
Reames, D. V. 1999, SSRv, 90, 413
Reames, D. V. 2009, ApJ, 693, 812
Reames, D. V., Barbier, L. M., & Ng, C. K. 1996, ApJ, 466, 473
Reames, D. V., Ng, C. K., & Tylka, A. J. 2013, SoPh, 285, 233
Rice, W. K. M., Zank, G. P., & Li, G. 2003, JGRA, 108, 1369
Rodríguez-Pacheco, J., Wimmer-Schweingruber, R. F., Mason, G. M., et al. 2020, A&A, 642, A7
Romeo, O. M., Braga, C. R., Badman, S. T., et al. 2023, ApJ, 954, 168
Sachdeva, N., Tóth, G., Manchester, W. B., et al. 2021, ApJ, 923, 176
Sachdeva, N., van der Holst, B., Manchester, W. B., et al. 2019, ApJ, 887, 83
Sarris, E. T., Anagnostopoulos, G. C., & Krimigis, S. M. 1987, JGRA, 92, 12083
Schatten, K. H., Wilcox, J. M., & Ness, N. F. 1969, SoPh, 6, 442
Sokolov, I. V., Holst, B. v. d., Manchester, W. B., et al. 2021, ApJ, 908, 172
Sokolov, I. V., van der Holst, B., Oran, R., et al. 2013, ApJ, 764, 23
Tóth, G., Sokolov, I. V., Gombosi, T. I., et al. 2005, JGRA, 110, A12226
Tóth, G., van der Holst, B., & Huang, Z. 2011, ApJ, 732, 102
Tóth, G., van der Holst, B., Sokolov, I. V., et al. 2012, JCoPh, 231, 870
Trotta, D., Larosa, A., Nicolaou, G., et al. 2024, ApJ, 962, 147
van der Holst, B., Sokolov, I. V., Meng, X., et al. 2014, ApJ, 782, 81
Vievering, J. T., Vourlidas, A., & Krucker, S. 2024, ApJ, 972, 48
Vourlidas, A., Howard, R. A., Plunkett, S. P., et al. 2016, SSRv, 204, 83
Wang, B. B., Zank, G. P., Zhao, L. L., & Adhikari, L. 2022, ApJ, 932, 65
Wiedenbeck, M. E., Angold, N. G., Birdwell, B., et al. 2017, ICRC, 35, 16
Wijsen, N., Lario, D., Sánchez-Cano, B., et al. 2023, ApJ, 950, 172
Wraback, E., Hoffmann, A., Manchester, W., et al. 2024, ApJ, 962, 182
Xu, Z., Cohen, C., Cummings, A., et al. 2024, AGU Fall Meeting, 2024, SH33C–2743
Zank, G. P., Rice, W. K. M., & Wu, C. C. 2000, JGR, 105, 25079
Zhao, L., Sokolov, I., Gombosi, T., et al. 2024, SpWea, 22, e2023SW003729